\documentclass[12pt,preprint]{article}
\usepackage{amsmath}
\usepackage{graphics}
\usepackage{epsfig}
\renewcommand{\vec}[1]{{\bf #1}}
\newcommand{\eqb}{\begin{equation}}
\newcommand{\eqe}{\end{equation}}
\newcommand{\dmb}{\begin{displaymath}}
\newcommand{\dme}{\end{displaymath}}
\newcommand{\pd}{\partial}

\newcommand{\eab}{\begin{eqnarray}}
\newcommand{\eae}{\end{eqnarray}}
\newcommand{\ra}{\right\rangle}
\newcommand{\la}{\left\langle}

\newcommand{\be}{\begin{equation}}
\newcommand{\ee}{\end{equation}}

\newcommand{\La}{\Lambda}
\begin{document}
\begin{titlepage}
\begin{flushright} 
\end{flushright}
\vspace{0.6cm}

\begin{center}
\Large{Yang-Mills thermodynamics: The confining phase}

\vspace{1.5cm}

\large{Ralf Hofmann}

\end{center}
\vspace{1.5cm}

\begin{center}
{\em Institut f\"ur Theoretische Physik\\ 
Universit\"at Frankfurt\\ 
Johann Wolfgang Goethe - Universit\"at\\ 
Robert-Mayer-Str. 10\\ 
60054 Frankfurt, Germany}
\end{center}
\vspace{1.5cm}

\begin{abstract}
We summarize recent nonperturbative results obtained for the 
thermodynamics of an SU(2) and an SU(3)
Yang-Mills theory being in its 
confining (center) phase. This phase is associated with a 
dynamical breaking of the local magnetic 
center symmetry. Emphasis is 
put on an explanation of the involved 
concepts.   

\end{abstract} 

\end{titlepage}

\noindent{\sl Introduction.} This is the last one in a series of three papers giving an 
abbreviated presentation of nonperturbative concepts and results for the 
thermodynamics of an SU(2) or an SU(3) Yang-Mills theory as obtained in 
\cite{Hofmann2005,HerbstHofmannRohrer2004,HerbstHofmann2004}. Here we discuss the 
confining or center phase. 

The three unexpected results for the confining phase are the spin-1/2 
nature of the massless (neutral, Majorana) and massive (charged) excitations, the Hagedorn 
nature of the transition from the confining to the preconfining phase, 
and the exact vanishing of the pressure and energy density of the 
ground state in the confining phase of an SU(2) or SU(3) 
Yang-Mills theory. 

The first result clashes 
with the perception about 
bosonic glueballs being the observable excitations of 
pure SU(3) Yang-Mills theory at 
zero temperature. This statement seems to be supported by 
lattice simulations \cite{MorningstarPeardon1999} and 
by analysis based on the QCD-sum-rule method \cite{Narison1984}. 
We have discussed in \cite{Hofmann2005} why lattice 
simulations of pure SU(2) and SU(3) Yang-Mills theory 
run into a severe finite-size problem 
at low temperatures and thus are unreliable. QCD spectral 
sum rules \cite{ShifmanVainshteinZakharov1978}, on the other hand, {\sl assume} 
the existence of a lowest resonance with finite coupling 
to the currents of a given production 
channel (these currents are formulated as local functionals 
of the fundamental fields in the QCD Lagrangian). The resonance's properties are 
determined subsequently by assuming the analyticity of the associated correlation 
function in the external momentum and by appealing to an operator product expansion in the
deep euclidean region. Analyticity, however, must break 
down across two phase boundaries (deconfining-preconfining, preconfining-center) provided 
that the effects arising due to the deviation from the thermodynamical limit can, 
on a qualitative level, be neglected in the production process. As a consequence, 
the QCD sum rule method is probably unreliable for the 
investigation of the spectrum of a {\sl pure} SU(2) and SU(3) 
Yang-Mills theory. (We hasten to add that 
the situation is different for real-world hadronic resonances because the dynamical mixing 
of pure SU(3) and pure dual SU(3) gauge theories 
may restore a quasi-analytical behavior of the relevant 
correlation functions \cite{Hofmann2005}. After all the overwhelming 
phenomenological successes of QCD sum rules are a lot more more than 
coincidence. The term `dynamical mixing' includes the occurrence of 
the fractional Quantum Hall effect \cite{Laughlin1983,TsuiStoermerGossard1982} 
which renders quarks to be emerging phenomena, 
for a discussion see \cite{Hofmann2005}.)  

The second result -- the Hagedorn nature of the 
transition to the truly confining phase -- was suspected 
to occur for the real-world strong interactions 
a long time ago \cite{Hagedorn1965,Veneziano1968}. Subsequently performed lattice 
simulations seemed to exclude a Hagedorn transition 
(diverging partition function above the critical point) even in the case of a 
pure SU(2) or SU(3) Yang-Mills theory \cite{LuciniTeperWenger}. Again, this a 
consequence of the lattice's failure to properly 
capture the infrared physics in 
thermodynamical simulations at low temperature, for an 
extended discussion see \cite{Hofmann2005}. 

The third result, namely the excact vanishing of the ground-state pressure 
and energy density of the Yang-Mills theory at zero temperature, commonly is 
used as a normalization assumption in lattice computations \cite{Bielfeld19961999} 
and not obtained as a dynamical result. In \cite{Hofmann2005} we have shown the absoluteness, 
that is, the {\sl gravitational} measurablility of the finite and exactly 
computable energy-momentum tensor associated with the 
ground-state in the deconfining and preconfining phases: An immediate consequence of 
the fact that these ground states are determined by 
radiatively protected BPS equations. (Since these equations are first-order as 
opposed to second-order Euler-Lagrange equations the usual shift 
ambiguity in the corresponding potentials is absent.) Recall, that the 
finiteness of the ground-state energy density and 
pressure in the deconfining and preconfining phases arises from averaged-over interactions 
between and radiative corrections within solitonic field configurations. 
Being (euclidean) BPS saturated, classical configurations in the deconfining phase the 
latter are free of pressure and energy density in isolation. The same applies 
to the massless, interacting magnetic monopoles which, by their condensation, form the ground state in 
the preconfining phase. In the confining phase configurations that are 
free of pressure and energy density do also exist (single center-vortex loops). In contrast to the 
other phases propagating gauge field fluctuations are, however, absent in the confining phase. 
Only contact interactions occur between the center solitons, which, however, do not elevate the vanishing 
energy density of the isolated soliton to a finite value for the 
ensemble. The proof for this relies on computing the curvature 
of the potential for the spatial coarse-grainined center-vortex 
condensate at its zeros and by comparing this curvature with the square of the maximal 
resolution that is allowed for in the 
effective theory \cite{Hofmann2005}, see below. 

The outline of this paper is as follows: First, we 
discuss the occurence of isolated, instable, that is, contracting and 
collapsing center-vortex loops in the preconfining phase. 
From the evolution of the magnetic coupling constant in this 
phase we conclude that center-vortex loops become stable, particle-like 
excitations at the deconfining-confining phase boundary. 
Second, we point out the spin-1/2 nature of these particles, and we derive a dimensionless 
parameter with discrete values describing the condensate 
of pairs of single center-vortex loops 
after spatial coarse-graining. A discussion of 
the creation of center-fluxes (local phase jumps of the vortex condensate) 
by the decay of the monopole 
condensate in the preconfining phase is given. Third, we 
construct potentials for the vortex condensates which, in their physical effects, 
are uniquely determined by the remaining local symmetry and by the positive 
semi-definiteness of the energy density: Particle creation 
by local phase jumps of the order parameter 
may only go on so long as the energy density feeding into their 
creation is nonvanishing. Fourth, we discuss in detail 
the remarkable result that for SU(2) and SU(3) Yang-Mills dynamics 
the confining phase's ground-state energy density is exactly nil. 
In particular, we stress the fact that radiative 
corrections to the tree-level result are 
entirely absent. Fifth, we give an estimate for the density of 
static fermion states and thus establish the Hagedorn nature of 
the transition from the confining to the preconfining phase. Finally, 
we summarize our results in view of its implications 
for particle physics and for cosmology.\vspace{0.3cm}\\ 
\noindent{\sl Instable center-vortex loops in the preconfining phase.} 
Here we discuss the SU(2) case only, results for SU(3) follow by 
simple doubling. The ground-state of the preconfining 
phase is a condensate of magnetic monopoles peppered with 
instable defects: closed magnetic flux lines whose core 
regions dissolve the condensate locally and thus restore the 
dual gauge symmetry U(1)$_D$ (for SU(3): U(1)$^2_D$). 
It was shown in \cite{Hofmann2005} that the magnetic 
flux carried by a given vortex-loop solely depends on 
the {\sl charge} of the monopoles and antimonopoles contributing 
to the explicit magnetic current inside the vortex core. Thus the various species 
of vortex-loops, indeed, are mapped one-to-one 
onto the nontrivial center elements of SU(2) or SU(3): They deserve the 
name center-vortex loops. In the magnetic phase, center-vortex 
loops are, however, instable as we show now. To derive the classical 
field configuration associated with an infinitely 
long vortex line one considers an Abelian Higgs model 
with no potential and a magnetic coupling $g$. 
(We need to discuss the energy-momentum tensor 
of the solitonic configuration {\sl relative} to the ground state 
obtained by spatially averaging over instable vortex loops. Thus 
we need to substract the temperature dependent 
ground-state contribution which is reached far away from the considered vortex core 
as a result of the applicable spatial coarse-graining, 
see \cite{Hofmann2005} for details.) The following 
ansatz is made for the static dual gauge field $a^D_\mu$ \cite{NielsenOlesen1973}:
\eqb
\label{ANOansatz}
a_4^D=0\,,\ \ \ \ \ \ \ \ \ a_i^D=\epsilon_{ijk}\hat{r}_j e_k\, A(r)
\eqe
where $\hat{r}$ is a radial unit vector in 
the $x_1x_2$ plane, $r$ is the distance from the vortex core, 
and $\vec{e}$ denotes a unit vector along the vortex' symmetry axis which we choose to coincide with 
the $x_3$ coordinate axis. No analytical 
solution with a finite energy per vortex length 
is known to the system of the two coupled equations of motion 
honouring the ansatz (\ref{ANOansatz}) and the Higgs-field 
decomposition $\varphi=|\varphi|(r)\exp[i\theta]$. An approximate 
solution, which assumes the constancy of $|\varphi|$, is given as
\eqb
\label{A(r)}
A(r)=\frac{1}{gr}-|\varphi|K_1(g|\varphi|r)\longrightarrow \frac{1}{gr}-
|\varphi|\sqrt{\frac{\pi}{2g|\varphi|r}}\exp[-g|\varphi|r]\,,\ \ \ \ \ (r\to\infty)\,.
\eqe
In Eq.\,(\ref{A(r)}) $K_1$ is a modified Bessel function. Outside the core region 
the isotropic pressure $P_v(r)$ in the $x_1x_2$ plane is, up to an exponentially 
small correction, given as
\eqb
\label{vortexpres}
P_v(r)=-\frac{1}{2}\frac{\Lambda_M^3\beta}{2\pi}\,\frac{1}{g^2 r^2}\,.
\eqe
Notice that we have substituted the asymptotic value 
$|\varphi|=\sqrt{\frac{\Lambda_M^3\beta}{2\pi}}\,,\ (\beta\equiv\frac{1}{T})\,,$ 
as it follows from the spatially coarse-grained 
action in the preconfining (or magnetic) phase \cite{Hofmann2005,Hofmann20052}. Notice also the minus 
sign on the right-hand side of Eq.\,(\ref{vortexpres}): The 
configuration in Eq.\,(\ref{A(r)}) is static due to its cylindrical symmetry 
but highly instable w.r.t. bending of the vortex axis. In particular, 
the pressure inside a center-vortex {\sl loop} is more negative than outside causing the soliton to 
contract, and, eventually, to dissolve. Bending of the vortex axis occurs because 
there are no isolated magnetic charges in the preconfining phase which could serve as sources 
for the magnetic flux. An equilibrium between vortex-loop 
creation by the spatially and temporally {\sl correlated} dissociation of large-holonomy calorons 
and vortex-loop collapse is responsible for the 
negative pressure of the ground state in the preconfining phase. 
The typical core-size $R$ of a center-vortex loop evaluates as 
$R\sim\frac{1}{m_{a^{\tiny D}_\mu}}=\frac{1}{g}\sqrt{\frac{\Lambda_M^3}{\beta}}$ and 
its energy as $E_v\sim\frac{\pi}{g}\sqrt{\frac{\Lambda_M^3\beta}{2\pi}}$. (This takes 
into account an estimate for $\varphi$'s gradient contribution to the total 
energy of the soliton.)  

Notice that core-size $R$, 
energy $E_v$, and pressure $P_v(r)$ of a center-vortex vanish 
in the limit $g\to\infty$. This situation 
is reached at the critical temperature $T_{c,M}$ where the magnetic 
coupling diverges in a logarithmic fashion: $g\sim-\log(T-T_{c,M})$ \cite{Hofmann2005}. 
At $T_{c,M}$ the creation of single center-vortex loops at rest with respect to 
the heat bath (i) does not cost any energy 
and (ii) entails the existence of stable and massless particles. The latter do, in turn, 
condense pairwise into a new ground state.\vspace{0.3cm}\\  
\noindent{\sl Pairwise condensation of 
single center-vortex loops: Ground-state decay and 
change of statistics.} We consider a static, 
circular contour $C(\vec{x})$ of infinite radius -- an $S_1$ -- 
which is centered at the point $\vec{x}$. In addition, at finite coupling $g$ 
we consider a system $S$ of two single center-vortex loops, 1 and 2, which both 
are pierced by $C(\vec{x})$ and which contribute opposite units of 
center flux $F_{v_1}=\frac{2\pi}{g}=\oint_{C(\vec{x})} dz_\mu\, a^D_{1,\mu}=-F_{v_2}$ 
through the minimal surface spanned by $C(\vec{x})$. Depending on 
whether 1 collapses before or after 2 or whether 1 moves away 
from $C$ before or after 2 the total center flux $F$ through $C'$s 
minimal surface reads
\eqb
\label{fluxiso}
F=\left\{\begin{array}{c}
\hspace{-3cm}\pm\frac{2\pi}{g}\ \ \ \ \ \ (\mbox{either 1 or 2 is pierced by $C(\vec{x})$})\\ 
0\ \ \ \ \ \ \ \ (\mbox{1 and 2 or neither 1 nor 2 are pierced by $C(\vec{x})$})\,.\end{array}\right.
\eqe
The limit $g\to\infty$, which dynamically takes place at $T_{c,M}$, causes
the center flux of the isolated system $S$ to vanish and renders single center-vortex 
loops massless and stable particles. The center flux of the isolated system $S$ 
does no longer vanish if we couple $S$ to the heat bath. Although 1 and 
2 individually are spin-1/2 fermions the system $S$ obeys 
bosonic statistics. (Both, 1 and 2, come in two 
polarizations: the projection of the dipole moment, generated by 
the monopole current inside the core of the center-vortex loop, onto a 
given direction in space either is parallel or 
antiparallel to this direction.) Thus, assuming the spatial momentum 
of 1 and 2 to vanish, the quantum statistical average flux reads
\eab
\label{avfluxsysV}
\lim_{g\to\infty}\,F_{\tiny\mbox{th}}&=&4\pi F\,\int d^3p\,\,\delta^{(3)}(\vec{p})\, n_B(\beta |2\,E_v(\vec{p})|)\nonumber\\ 
&=&0,\pm\frac{8\pi}{\beta|\varphi|}=0,\pm 4\,\lambda^{3/2}_{c,M}\,.
\eae
According to Eq.\,(\ref{avfluxsysV}) there are finite, discrete, and dimensionless 
parameter values for the description of the 
macroscopic phase 
\eqb
\label{phasePhi}
\Gamma\frac{\Phi}{|\Phi|}(\vec{x})\equiv \lim_{g\to\infty}\exp[i\la \oint_{C(\vec{x})}dz_\mu\, 
a^D_\mu\ra]
\eqe
associated with the Bose condensate of the system $S$. 
In Eq.\,(\ref{phasePhi}) $\Gamma$ is an undetermined and dimensionless 
complex constant. Notice that taking the limit of vanishing spatial momentum for 
each single center-vortex loop {\sl is} the implementation of 
spatial coarse-graining. This coarse-graining is performed 
down to a resolution $|\Phi|$ which is determined by the (existing) stable solution to the 
equation of motion in the effective theory, see below.  

\begin{figure}
\begin{center}
\leavevmode
\leavevmode
\vspace{4.5cm}
\includegraphics{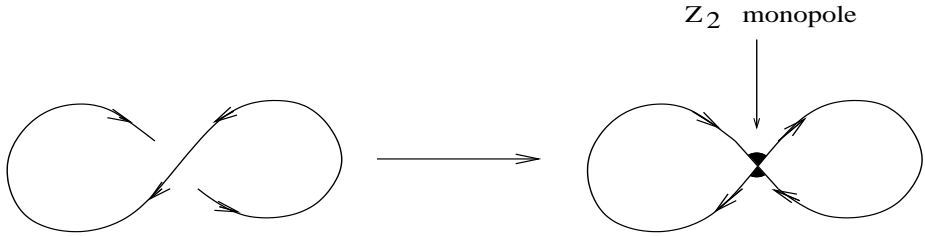}
\end{center}
\caption{The creation of an isolated $Z_2$ 
monopole by self-intersection of a 
center-vortex loop. \label{Fig-0}}      
\end{figure}

For convenience we normalize the parameter 
values given by  $\lim_{g\to\infty}\,F_{\tiny\mbox{th}}$
as $\hat{\tau}\equiv \pm 1,0$.\vspace{0.3cm}\\ 
\noindent{\sl Coarse-grained action and center jumps.}
To investigate the decay of the monopole condensate at $T_{c,M}$ (pre- and reheating) 
and the subsequently emerging equilibrium situation, we need to find conditions to constrain 
the potential $V_C$ for the macroscopic field $\Phi$ in such a way 
that the dynamics arising from it is unique. Recall that 
at $T_{c,M}$ the dual gauge modes of the preconfining phase 
decouple. Thus the entire process of fermionic pre- and reheating in the confining phase 
is described by spatially and temporally discontinuous changes of the 
modulus (energy loss) and phase (flux creation) of 
the field $\Phi$. Since the condensation of 
the system $S$ renders the expectation of the 't Hooft loop finite (proportional 
to $\Phi$) the magnetic center symmetry $Z_2$ (SU(2)) and $Z_3$ (SU(3)) 
is dynamically broken as a discrete gauge symmetry. Thus, 
after return to equilibrium, the ground state of the confining phase 
must exhibit $Z_2$ (SU(2)) and $Z_3$ (SU(3)) degeneracy. This implies that for SU(2) 
the two parameter values $\hat{\tau}=\pm 1$ need to be identified while each of the 
three values $\hat{\tau}=\pm 1,0$ describe a distinct 
ground state for SU(3). Let us now discuss how either one of 
these degenerate ground states is reached. Spin-1/2 particle creation 
proceeds by single center vortex loops being sucked-in from 
infinity. (The overall pressure is still negative during the 
decay of the monopole condensate thus facilitating the in-flow 
of spin-1/2 particles from spatial infinity.) At a given point $\vec{x}$ an observer 
detects the in-flow of a massless fermion in terms of the field 
$\Phi(\vec{x})$ rapidly changing its phase by a forward center jump 
(center-vortex loop gets pierced by $C(\vec{x})$) which is followed by the associated 
backward center jump (center-vortex loop lies inside $C(\vec{x})$). 
Each phase change corresponds to a tunneling 
transition inbetween regions of positive curvature in 
$V_C$. If a phase jump has 
taken place such that the subsequent potential 
energy for the field $\Phi$ is still positive then 
$\Phi$'s phase needs to perform additional jumps in order to 
shake off $\Phi$'s energy completely. This can only 
happen if no local minimum exists at a finite value of $V_C$. 
If the created single center-vortex loop moves sufficiently fast it can subsequently 
convert some of its kinetic energy into mass by twisting: massive, self-intersecting 
center-vortex loops arise. These particles are also spin-1/2 
fermions: A $Z_2$ or $Z_3$ monopole, constituting the 
intersection point, reverses the center 
flux \cite{Reinhardt2001}, see Fig.\,\ref{Fig-0}. 

If the SU(2) (or SU(3)) pure 
gauge theory does not mix with any other preconfining or deconfining 
gauge theory, whose propagating gauge modes would couple 
to the $Z_2$ (or $Z_3$) charges, a soliton generated 
by $n$-fold twisting is stable in isolation and possesses a 
mass $n\,\Lambda_C$. Here $\Lambda_C$ is the mass of the 
charge-one state (one self-intersection). After a sufficiently large and even number 
of center jumps has occurred the 
field $\Phi(\vec{x})$ settles in one of 
its minima of zero energy density. Forward - and backward tunneling inbetween 
these minima corresponds to the spontaneous on-shell generation of a massless, 
single center-vortex loop of zero momentum. In a WKB-like approximation one 
expects the associated euclidean trajectory to have a 
large action which, in turn, predicts large suppression. We conclude that tunneling 
between the minima of zero energy density 
is forbidden.  

\begin{figure}
\begin{center}
\leavevmode
\leavevmode
\vspace{5.5cm}
\includegraphics{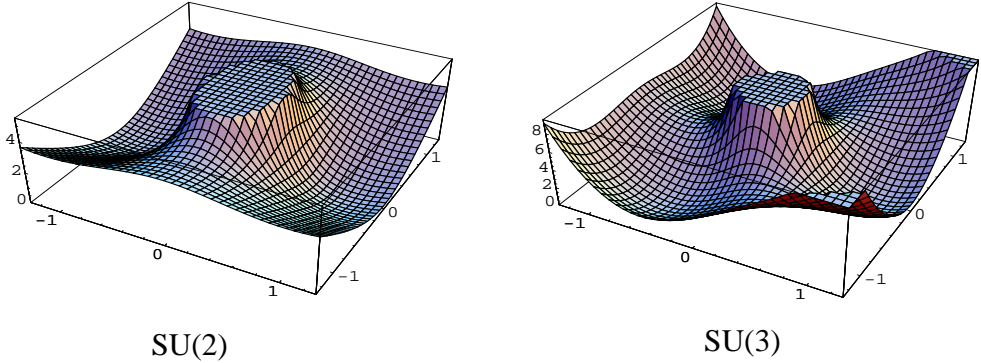}
\end{center}
\caption{The potential $V_C=\overline{v_C(\Phi)}v_C(\Phi)$ for the center-vortex 
condensate $\Phi$. Notice the regions of negative tangential curvature 
inbetween the minima.\label{Fig-1}}      
\end{figure}

Let us summarize the results of our above discussion: (i) the potential 
$V_C$ must be invariant under magnetic 
center jumps $\Phi\rightarrow\exp[\pi i]\Phi$\,\,(SU(2)) and 
$\Phi\rightarrow\exp[\pm \frac{2\pi}{3}]\Phi$\,\,(SU(3)) 
only. (An invariance under a larger continuous or 
discontinuous symmetry is excluded.) (ii) 
Fermions are created by a forward - and backward 
tunneling corresponding to local center jumps in $\Phi$'s 
phase. (iii) The minima of 
$V_C$ need to be at zero-energy density and are 
all related to each other by center transformations, 
no additional minima exist. (iv) 
Moreover, we insist on the occurrence of one mass 
scale $\Lambda_C$ only to parameterize the potential $V_C$. (As it was 
the case for the ground-state physics in the de - 
and preconfining phases.) (v) In addition, it is clear that the potential $V_C$ needs to 
be real.\vspace{0.1cm}\\           
\noindent\underline{SU(2) case:}\vspace{0.1cm}\\ 
A generic potential $V_C$ satisfying (i),(ii), (iii), (iv), and (v) is given by
\eqb
\label{2potC}
V_C=\overline{v_C}\,v_C\equiv\overline{\left(\frac{\Lambda_C^3}{\Phi}-\La_C\,\Phi\right)}\,
\left(\frac{\Lambda_C^3}{\Phi}-\La_C\,\Phi\right)\,.
\eqe
The zero-energy minima of $V_C$ 
are at $\Phi=\pm \Lambda_C$. It is clear that 
adding or subtracting powers $(\Phi^{-1})^{2l+1}$ or 
$\Phi^{2k+1}$ in $v_C$, where $k,l=1,2,3,\cdots$, 
generates additional zero-energy 
minima, some of which are {\sl not} related by center transformations 
(violation of requirement (iii)). Adding $\Delta V_C$, defined as 
an {\sl even}  power of a Laurent expansion 
in $\bar{\Phi}\Phi$, to $V_C$ (requirements (iii) and (v)), does in general 
destroy property (iii). A possible exception is 
\eqb
\label{DeltaVC}
\Delta V_C=\lambda\left(\Lambda_C^2-
\Lambda_C^{-2(n-1)}\left(\bar{\Phi}\Phi\right)^n\right)^{2k}
\eqe
where $\lambda>0; k=1,2,3,\cdots; n\in {\bf Z}$. Such a term, however, 
is irrelevant for the description of the tunneling 
processes (requirement (ii)) since the associated euclidean 
trajectories are essentially along U(1) Goldstone 
directions for $\Delta V_C$ due to the 
pole in Eq.\,(\ref{2potC}). Thus adding $\Delta V_C$ does not cost 
much additional euclidean action and therefore does not affect 
the tunneling amplitude in a significant way. As for the 
curvature of the potential at its minima, adding $\Delta V_C$ does not lower 
the value as obtained for $V_C$ alone. One 
may think of multiplying $V_C$ with a positive, dimensionless 
polynomial in $\Lambda_C^{-2}\bar{\Phi}\Phi$ with 
coefficients of order unity. This, however, 
does not alter the physics of the pre - and reheating 
process. It increases the curvature of the 
potential at its zeros and therefore does not alter the result that quantum 
fluctuations of $\Phi$ are absent after relaxation.   
\vspace{0.1cm}\\   
\noindent\underline{SU(3) case:}\vspace{0.1cm}\\ 
A generic potential $V_C$ satisfying (i),(ii), (iii), (iv), and (v) is given by
\eqb
\label{3potC}
V_C=\overline{v_C}\,v_C\equiv\overline{\left(\frac{\Lambda_C^3}{\Phi}-\Phi^2\right)}\,
\left(\frac{\Lambda_C^3}{\Phi}-\Phi^2\right)\,.
\eqe
The zero-energy minima of $V_C$ 
are at $\Phi=\Lambda_C\exp\left[\pm\frac{2\pi i}{3}\right]$ and $\Phi=\Lambda_C$. 
Again, adding or subtracting powers $({\Phi}^{-1})^{3l+1}$ or $(\Phi)^{3k-1}$ 
in $v_C$, where $l=1,2,3,\cdots$ and $k=2,3,4,\cdots$, 
violates requirement (iii). The same discussion for adding 
$\Delta V_C$ to $V_C$ and for 
multiplicatively modifying $V_C$ applies as in the SU(2) case. 
In Fig.\,\ref{Fig-1} plots of the potentials in Eq.\,(\ref{2potC}) and Eq.\,(\ref{3potC}) are shown. 
The ridges of negative tangential curvature are classically forbidden: 
The field $\Phi$ tunnels through these ridges, and a 
phase change, which is determined by an element of the center $Z_2$ (SU(2)) or $Z_3$ (SU(3)), 
occurs locally in space. This is the afore-mentioned generation of 
one unit of center flux.\vspace{0.3cm}\\ 
\noindent{\sl No vacuum energy after relaxation.}
The action describing the process of relaxation of $\Phi$ 
to one of $V_C$'s minima is  
\eqb    
\label{actionPhi}
S = \int d^4x 
\left(\frac{1}{2}\,\overline{\partial_\mu \Phi} 
\partial^\mu \Phi - \frac12\, V_C \right) \,.
\eqe
Once $\Phi$ has settled into $V_C$'s 
minima $\Phi_{\tiny\mbox{min}}$ there are no 
quantum fluctuations $\delta\Phi$ to 
be integrated out anymore. Let us show this: 
Writing $\Phi=|\Phi|\exp\left[i\frac{\theta}{\La_c}\right]$, we have  
\eqb
\label{minimacur}
\left.\frac{\pd^2_{\theta} V_C(\Phi)}{|\Phi|^2}\right|_{\Phi_{\tiny\mbox{min}}}=
\left.\frac{\pd^2_{|\Phi|} V_C(\Phi)}
{|\Phi|^2}\right|_{\Phi_{\tiny\mbox{min}}}
=\left\{\begin{array}{c}8\,\ \ \ \ \ (\mbox{SU(2)})\\ 
18\,\ \ \ \ \ (\mbox{SU(3)})\end{array}\right.\,.
\eqe
Thus a potential fluctuation $\delta\Phi$ would 
be harder than the maximal resolution $|\Phi_{\tiny\mbox{min}}|$ corresponding to 
the effective action Eq.\,(\ref{actionPhi}) that arises after 
spatial coarse-graining. Thus quantum fluctuations are already 
contained in the classical configuration $\Phi_{\tiny\mbox{min}}$: The cosmological constant 
in the confining phase of an SU(2) or 
SU(3) Yang-Mills theory vanishes exactly. Again, adding the term 
$\Delta V_C$ of Eq.\,(\ref{DeltaVC}) to the potentials in Eqs.\,(\ref{2potC}) and 
(\ref{3potC}) or performing the above multiplicative modification 
does not lower the value for the curvature as 
obtained in Eq.\,(\ref{minimacur}) and therefore 
does not change this result.\vspace{0.3cm}\\ 
\noindent{\sl Estimate for density of states, Hagedorn nature of the transition.} 
That the transition from the confining to the preconfining phase 
is of the Hagedorn nature is shown by an estimate for 
the density of massive spin-1/2 states. 
The multiplicity of massive fermion states, associated with center-vortex loops possessing 
$n$ self-intersections, is given by twice the number $L_n$
of bubble diagrams with $n$ vertices in a scalar $\lambda \phi^4$ theory. 
In \cite{BenderWu1969} the minimal number of such diagrams $L_{n,min}$ was 
estimated to be 
\eqb
\label{NOD}
L_{n,min}=n!3^{-n}\,.
\eqe
The mass spectrum is equidistant. That is, 
the mass $m_n$ of a state with $n$ self-intersections of 
the center-vortex loop is $m_n\sim n\,\La_C$. If we only 
ask for an estimate of the density of {\sl static} 
fermion states $\rho_{n,0}=\tilde{\rho}(E=n\La_C)$ of energy 
$E$ then, by appealing to Eq.\,(\ref{NOD}) and Stirling's formula, 
we obtain \cite{Hofmann2005}
\eab
\label{statdes}
\rho_{n,0}&>&\frac{\sqrt{8\pi}}{3\La_C}\,\exp[n\log n]\Big(\log n+1\Big)\,\ \ \ \ \ \mbox{or}\nonumber\\ 
\tilde{\rho}(E)&>&\frac{\sqrt{8\pi}}{3\La_C}\exp[\frac{E}{\La_C}\log\frac{E}{\La_C}]
\Big(\log\frac{E}{\La_C}+1\Big)\,.
\eae
Eq.\,(\ref{statdes}) tells us that the density of static fermion 
states is more than exponentially increasing with energy $E$. The partition function 
$Z_{\Phi}$ for the system of static fermions thus is estimated as
\eab
\label{partfunctionphi}
Z_{\Phi}&>&\int_{E^*}^\infty dE\,\tilde{\rho}(E)\,n_F(\beta E)\nonumber\\ 
&>&\frac{\sqrt{8\pi}}{3\La_C}\,\int_{E^*}^\infty dE\,\exp\left[\frac{E}{\La_C}\right]\,
\exp[-\beta E]\,,
\eae
where $E^*\gg \Lambda_C$ is the energy where we start to trust 
our approximations. Thus $Z_{\Phi}$ diverges at some 
temperature $T_H<\Lambda_C$. Due to the logarithmic factor in the exponent 
arising in estimate Eq.\,(\ref{statdes}) for $\tilde{\rho}(E)$ we 
would naively conclude that $T_H=0$. This, however, is an artefact 
of our assumption that all states with $n$ self-intersections 
are infinitely narrow. Due to the existence of contact interactions between 
vortex lines and intersection points this assumption is 
the less reliable the higher the total energy of a 
given fluctuation. (A fluctuation of large energy has a higher 
density of intersection points and vortex lines and 
thus a larger likelihood for the occurrence of contact interactions which mediate the 
decay or the recombination of a given state with $n$ self-intersections.) 
At the temperature $T_H$ the entropy 
wins over the Boltzmann suppression in energy, and the partition 
function diverges. To reach the point $T_H$ one would, in a spatially homogeneous way, 
need to invest an infinite amount of energy into the system 
which is impossible. By an (externally induced) violation of 
spatial homogeneity and thus by a sacrifice of thermal 
equilibrium the system may, however, condense densly packed (massless) 
vortex intersection points into a new ground state. The latter's excitations exhibit a 
power-like density of states and thus are described by a 
finite partition function. This is the celebrated Hagedorn 
transition from below.\vspace{0.3cm}\\ 
\noindent{\sl Summary in view of particle physics and cosmology.} The confining 
phase of an SU(2) and SU(3) pure Yang-Mills theory is characterized 
by a condensate of single center-vortex loops and a dynamically broken, local magnetic 
$Z_2$ (SU(2)) and $Z_3$ (SU(3)) symmetry: No massless or finite-mass gauge bosons exist. 
Single center-vortex loops 
emerge as massless spin-1/2 particles due to 
the decay of a monopole condensate. A fraction of zero-momentum, 
single center-vortex loops subsequently condenses by the formation 
of Cooper-like pairs. Protected from radiative corrections, the 
energy density and the pressure of this condensate is 
precisely zero in a thermally equilibrated situation. The spectrum of particle 
excitations is a tower of spin-1/2 states with 
equidistant mass levels. A massive state emerges by twisting 
a single center-vortex loop hence generating 
self-intersection point(s). This takes place when 
single center-vortex loops collide. The process of mass generation thus 
is facilitated by converting (some of) the kinetic energy of a single center-vortex loop 
into the (unresolvable) dynamics of a flux-eddy 
marking the self-intersection point, see Fig.\,(\ref{Fig-0}). 
Due to their over-exponentially increasing multiplicity 
heavy states become instable by the contact interactions facilitated by 
dense packing. In a spatially extended 
system (such as the overlap region for two colliding, heavy, and 
ultrarelativistic nuclei) there is a finite value 
in temperature, comparable to the Yang-Mills scale $\Lambda_C$, 
where a given, spatially nonhomogeneous 
perturbation induces the condensation of vortex intersections. 
This is the celebrated (nonthermal) Hagedorn transition. 

The existence of a Hagedorn-like density of states explains 
why in an isolated system, governed by a single SU(2) Yang-Mills theory, 
the center-flux eddy in a spin-1/2 state with a single self-intersection appears to be 
structureless for external probes of all momenta 
with one exception: If the externally supplied resolution is 
comparable to the Yang-Mills scale $\Lambda_C$, that is, close 
to the first radial excitation level of a BPS monopole 
\cite{ForgasVolkov2003} then the possibility of converting the 
invested energy into the entropy associated with 
the excitation of a large number of instable and 
heavy resonances does not yet exist. As a consequence, the center 
of the flux eddy  -- a BPS monopole -- is excited itself and 
therefore reveals part of its structure. For an externally supplied 
resolution, which is considerable below $\Lambda_C$, there is nothing to 
be excited in a BPS monopole \cite{ForgasVolkov2003} and thus the object 
appears to be structureless as well. 

There is experminental evidence \cite{Alvensleben1968,Ashkin1953,Scott1951} that this situation 
applies to charged leptons being the spin-1/2 states with a single self-intersection of 
SU(2) Yang-Mills theories with scales comparable with the associated 
lepton masses \cite{Hofmann2005}. The corresponding neutrinos are 
Majorana particles (single center-vortex loops) which is also supported 
by experiment \cite{Klapdor2004}. The weak symmetry SU(2)$_W$ of the Standard Model (SM) is 
identified with SU(2)$_e$ where the subscript $e$ refers to 
the electron. The important difference compared with the SM is 
that the pure SU(2)$_e$ gauge theory {\sl by itself provides} for a nonperturbative 
breakdown of its continuous gauge symmetry in 
two stages (deconfining and preconfining phase) 
and, in addition, generates the electron neutrino and the electron 
as the only stable and apparently structureless solitons in its confining 
phase: No additional, fundamentally charged, and fluctuating Higgs 
field is needed to break the weak gauge symmetry. The confining phase is associated with a 
discrete gauge symmetry -- the magnetic center symmetry -- being 
dynamically broken. 

As far as the cosmological-constant problem is concerned the state of affairs 
is not as clear-cut as it may seem. Even though each pure SU(2) or 
SU(3) gauge theory does not generate a contribution to the vacuum 
energy in its confining phase one needs to include 
gravity, the dynamical mixing of various 
gauge-symmetry factors, and the anomalies of 
emerging global symmetries in the analysis 
to obtain the complete picture on 
the Universe's present ground state. We hope to be able to pursue 
this program in the near future. Notice that today's ground-state 
contribution due to an SU(2) Yang-Mills 
theory of scale comparable to the present temperature of the 
cosmic microwave background is small as 
compared to the measured value \cite{Hofmann20054}. This SU(2) Yang-Mills 
theory masquerades as the U(1)$_Y$ factor of the Standard Model 
within the present cosmological epoch.

\section*{Acknowledgements}
The author would like to thank Mark Wise for useful conversations 
and for the warm hospitality extended to him during his 
visit to Caltech in May 2005. Financial support by 
Kavli Institute at Santa Barbara and by 
the physics department of UCLA is thankfully acknowledged.

\baselineskip25pt
\end{document}